\documentclass[aps,preprint,amsmath,amssymb,showpacs]{revtex4-1}
\usepackage{graphicx}

\newcommand{\nc}{NC}

\newcommand{\reffig}[1]{Fig.\ref{#1}}
\newcommand{\refeq}[1]{Eq.\ref{#1}}

\begin{document}

\title{ Theoretical analysis of electronic band structure of 2-to-3-nm Si nanocrystals }

\author{Prokop Hapala}
\email{hapala@fzu.cz}
\author{Kate\v{r}ina K\r{u}sov\'{a}}
\author{Ivan Pelant}
\author{Pavel Jel\'{i}nek}
\affiliation{Institute of Physics, Academy of Sciences of the Czech Republic, Cukrovarnick\' a 10, 162 00 Prague, Czech Republic}

\hyphenation{pa-ram-e-trized}

\pacs{73.21.-b, 73.21.La, 78.67.-n, 78.67.Hc }

\begin{abstract}
We introduce a~general method which allows reconstruction of electronic band structure of nanocrystals from ordinary real-space electronic structure calculations.  
A~comprehensive study of band structure of a~realistic nanocrystal is given including full geometric and electronic relaxation with the surface passivating groups. In particular, we combine this method 
with large scale density functional theory calculations to obtain insight into the luminescence properties of silicon nanocrystals of up to 3 nm in size depending on the surface passivation and geometric 
distortion. We conclude that the band structure concept is applicable to silicon nanocrystals with diameter larger than $\approx$ 2 nm  with certain limitations. 
We also show how perturbations due to polarized surface groups or geometric distortion can lead to considerable moderation of momentum space selection rules.
\end{abstract}

\maketitle

\section{ Introduction }

Crystalline nanostructures are often viewed as ``artificial atoms'' or ``zero-dimensional systems'' despite being the size of a smaller protein. 
This is because when single nanocrystals (\nc s) are experimentally probed, very sharp optical radiative transitions, implying the existence of discrete energy 
levels rather than energy bands, are observed.\cite{Brunner-FirstSingleNcSiSpectra-PRL1992} On the other hand, nanostructures are known to retain some of 
the band-structure-related properties of their bulk counterparts, e.g.\ direct-bandgap semiconductor \nc s such as CdSe are excellent light emitters.

This brings about an apparent dichotomy and raises the question if, and under what circumstances, the band structure concept can be applied to \nc s and, moreover, if the band structure description can overlap with the description based on discrete energy levels. In this article, we show that band structure can be rigorously defined in nanostructures, taking into account several alterations with respect to bulk materials. 

The concept of band structure in \nc s is very important, because it allows one to transfer the already thoroughly-studied electronic properties and the 
whole framework of solid-state physics of bulk materials to the nanoscale. It might not play such a crucial role in direct-bandgap materials, which have already 
been successfully theoretically described using models completely disregarding the band structure,\cite{Yoffe-SemiconductorNcReview} but it  starts to be of 
particular importance for the case of indirect-bandgap semiconductors. 

In indirect-bandgap semiconductors, the comprehension of the electronic band structure behavior allows for efficient band engineering. A~prime example illustrating the 
usefulness of the band-structure approach is germanium, an originally indirect-bandgap material successfully transformed to a direct-bandgap one by heavy doping and 
tensile strain.\cite{Sun-directGe-APL2009} This experimental realization of direct-gap bulk germanium was based on theoretical band-structure calculations.\cite{Liu-directGeTheory-OE2007} 
What is important, however, is that the same concept was experimentally realized also in germanium \nc s.\cite{Nataraj-DirectPLGeNc-OE2010}  
This example confirms that the parallels between bulk and nanocrystalline materials do exist and that they can be beneficial in material engineering as well as for the intuitive understanding of the material's behavior.

It is evident from the discussion above that in direct-bandgap materials only the states close to the valence and conduction band edges (situated at the $\Gamma$ point) will be those governing the luminescence behavior of the material. However, in indirect-bandgap semiconductors band structure can be influenced in a~more complex way, making it more difficult to decipher the resulting more complex behavior. This is why we chose to analyze the band structure of nanocrystalline silicon, as silicon is the most wide-spread indirect semiconductor.

Silicon as material found wide applications in electronics and it is still a subject of intense research. Nevertheless, the indirect band gap nature of the band structure of bulk silicon has always been the major obstacle for its employment in light-emitting devices since momentum conservation requires additional momentum transfer mechanisms involved in the light emission processes. The situation changed dramatically in the last two decades due to the emergence of the possibility of preparing Si-based structures with nanometer size, where quantum effects begin to play a dominant role.   
In particular, large effort has been devoted to the study of the optical properties of Si \nc s in the last years, with a~perspective of potential for real-life applications such as e.g.\ light emitting diodes, next-generation solar cells and biomedical devices. \cite{Khriachtchev2008, Kumar2008, Koshida2009, Pavesi2010} The discovery of efficient visible photoluminescence\cite{Canham-PorousSiPL-APL1990} (PL) and optical gain\cite{Pavesi-OpticalGain-N2000} from silicon \nc s has demonstrated the possibility of partially overcoming the limitations of the indirect band gap of silicon by exploiting the quantum phenomena at the nanoscale. Despite the large amount of papers  published on this subject, there are still  many aspects
which are not fully understood and are subject of intense dispute.

As already mentioned, the theoretical concept of optical properties of Si \nc s 
is often discussed in the framework of the band structure picture of the bulk material. However, the finite size of a~system measuring only a  few nanometers makes the justification of this approach questionable. Therefore, the validity of the band structure concept and, if need be, its character in Si \nc s of a given size and surface passivation have to be analyzed.  

Traditional electronic structure calculations of Si \nc s, as a~finite system, provide only real space molecular orbitals (MOs), where crystal momentum and thus band structure $E(\mathbf{k})$ is not directly accessible. What's more, in \nc s with typical sizes of a few nanometers, the surface-to-volume ratio substantially increases and thus the interface between the \nc s and their environment plays a crucial role in tailoring their optical properties. Consequently, the optoelectronic properties of Si \nc s are very sensitive to their surface passivation, symmetry and applied strain, which makes the application of simple (bulk like) models questionable.

To the best of  our knowledge, only a few attempts\cite{TraniPRB2005, PuschnigScience09, Valentin2008, Hu2002} have been made to perform the projection of states of finite systems from real to reciprocal space and  vice versa so far. Therefore, a~robust method which allows the band structure mapping from fully relaxed density functional theory (DFT) calculations of realistic Si \nc s is required.

From the perspective of theoretical simulations, many approaches, ranging from {\it ab initio}\cite{Guerra2009, Guerra2010, Konig2008, Konig2009, Seino2010} to parametrized semiempirical methods \cite{TraniPRB2005, Trani2007, Moskalenko2012, Poddubny2010} , have been adopted to investigate the optical and electronic properties of Si \nc s. {\it Ab initio} methods provide a very accurate description of electronic states of fully relaxed Si \nc s, but simulations of realistic systems with nanometer size are impossible due to excessive computational demand. On the other hand, semiempirical methods \cite{TraniPRB2005} might address Si \nc s of realistic size consisting of tens of thousands of atoms, but with limited transferability. From this point of view, the application of fast local orbital DFT codes \cite{Review-Fireball2011} devised with the aim of computational efficiency but still providing desired precision in description of ground state electronic structure seems to be the optimal choice.

In this paper, we will introduce a~general method which allows the reconstruction of band structure from MOs obtained e.g.\ from ab initio calculations. For this reason, we dare to use the terms highest occupied molecular orbitals (HOMO) and lowest unoccupied molecular orbital (LUMO) and valence-band maximum/conduction-band minimum as synonyms throughout this article. We will demonstrate the method's capability to fold up the band structure of finite systems on a~simple 1D atomic chain consisting of a few H atoms. Using this example, we will discuss the main characteristic features of the band structure of finite size systems. Next, we will apply the procedure to analyze the electronic structure of freestanding Si \nc s obtained from DFT calculations. To achieve this objective, we will adopt fast local orbital DFT code Fireball \cite{Jelinek2005PRB, Review-Fireball2011}, which allows us to perform fully relaxed total energy calculation of different Si \nc s consisting of up to thousand of atoms in a feasible 
way. We will show how the electronic band structure of Si \nc s is affected by their surface passivation, symmetry and size.

\section{ Methods }

\subsection{ DFT calculations }
All computations were carried out using the local orbital density functional theory (DFT) code FIREBALL \cite{Jelinek2005PRB, Review-Fireball2011, FireballDemkov1995} within the local-density approximation (LDA) for the exchange-correlation functional. Valence electrons have been described by optimized \cite{FireballDemkov1995} numerical atomic-like orbitals having the following cutoff radii (in a.u.): Rc(s, s*) = 4.0 for H; Rc(s) = 4.5, Rc(p) = 4.5, and Rc(d) = 5.4 for C; Rc(s) = 4.8, Rc(p) = 5.4, and Rc(d) = 5.2 for Si, and Rc(s) = 3.5, Rc(p) = 4.0, and Rc(d) = 5.0 for O, respectively. The correctness of the basis set was checked to reproduce the band structure of bulk silicon within LDA accuracy.    

\begin{figure*}[!!!t]
\centering
\includegraphics[scale=0.5]{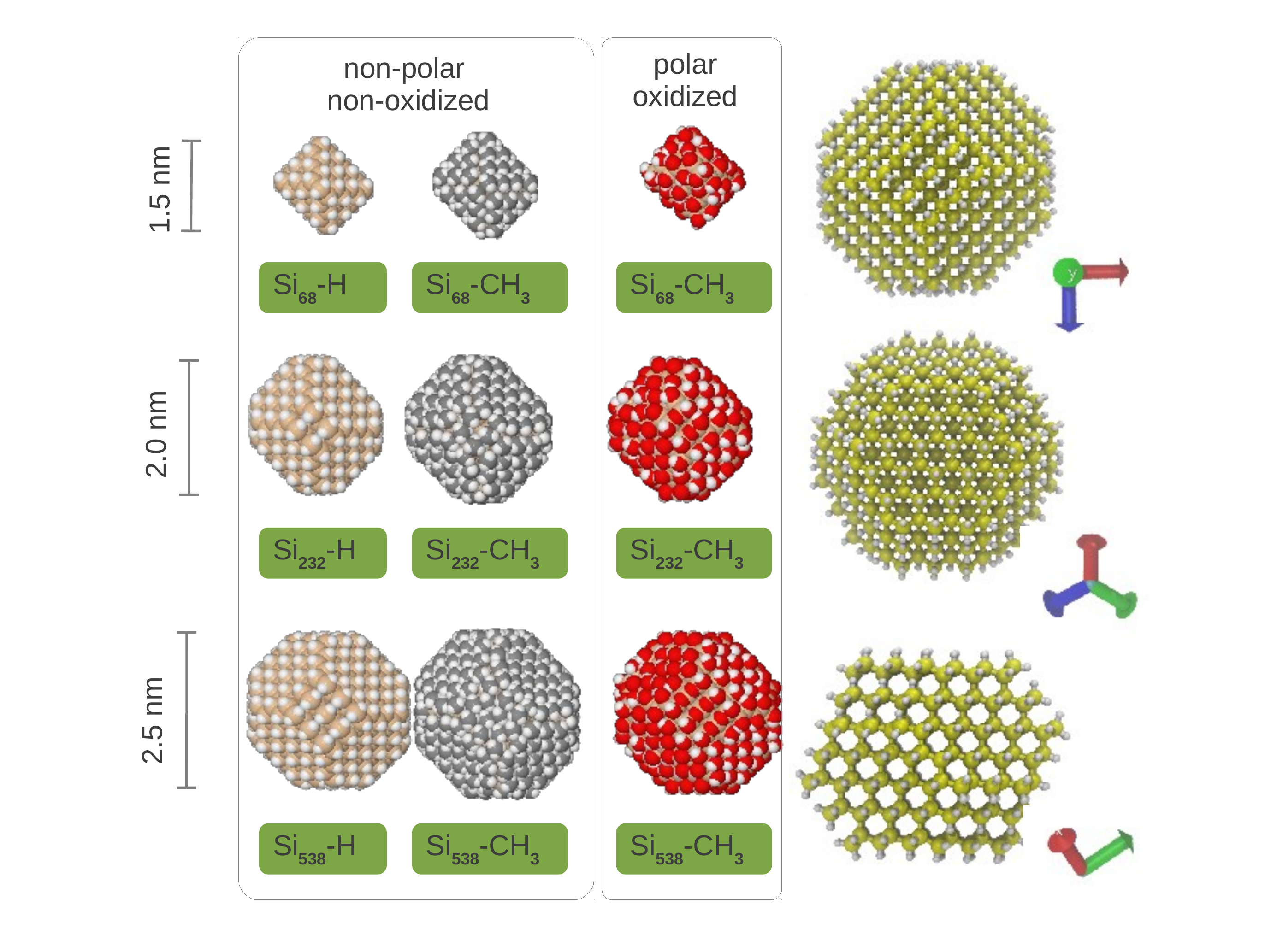}
\caption{  Geometry of model nanocrystals. First two panels show 9 different models of \nc s with 3 core sizes (68, 232 and 538 Si atoms of diameter 1.5, 2.0 and 2.5 nm) with either polar oxidized (--OH) or non-polar (--H, --CH$_3$) passivation. The last panel shows the geometry of Si$_{538}$-H from important crystallographic directions (100),(111) and perpendicularly to (111). }
\label{Geometry}
\end{figure*}

Nanocrystals are represented by cluster models with 3 different core sizes consisting of 68, 232 and 538 Si atoms, respectively.  These 3 models represent Si \nc s with the diameters of 1.5, 2.0 and 2.5 nm \reffig{Geometry}. 

Si core of each of the three models is terminated either with polar (--OH, simulating an oxidized layer) or non-polar (--H, --CH$_3$) passivating groups. Atomic structure of the Si$_{538}$ core was cut out from relaxed bulk silicon lattice in such a way so as to minimize the number of unsaturated bonds and to reflect the lattice symmetry with well-defined (111) and (100) faces ( \reffig{Geometry} ). Subsequently, the smaller Si$_{232}$ and Si$_{68}$ cores were derived from Si$_{538}$ by removing the topmost atomic layer, in order to get a smaller analog of the same symmetry and surface faces. Then, passivating groups of the three types were attached to the Si core, leaving all surface Si atoms fully saturated. 

All Si \nc s models were fully optimized, allowing the relaxation of all atoms. Total energy calculations were performed as a cluster calculation ( $\mathbf{k}$=0 ) and the convergence was achieved when the residual total energy of 0.0001 eV and the maximal force of 0.05 eV/ \r{A} were reached. 

\subsection{ Momentum space projection of molecular orbitals }

The band structure theory of solids is well established and it  has been successfully applied to real materials to explain their physical and material properties, e.g.\ electrical resistivity and optical absorption. A~corner-stone of the band structure theory is the so-called Bloch's theorem,\cite{Bloch1928} which had initiated the epoch of modern solid state physics. Electronic states in an infinite periodic system are described via the Schroedinger equation:

\begin{equation}
[\nabla^2 + V(\mathbf{r})]\Psi(\mathbf{r}) = \epsilon \Psi(\mathbf{r}),
\label{eqSch}
\end{equation}
where $\epsilon$ is the energy of eigenstate $\Psi(\mathbf{r})$ and potential $V(\mathbf{r}) $ is periodic $V(\mathbf{r}) = V(\mathbf{r + R})$, $\mathbf{R}$ being the translational lattice vector. Bloch's theorem postulates that the solution of the Schroedinger equation, $\Psi_{n,\mathbf{k}}(\mathbf{r}) $, can be written as a product of a~real mother-function $ u_{n,\mathbf{k}}(\mathbf{r}) $, which is also periodic $ u_{n,\mathbf{k}}(\mathbf{r}) = u_{n,\mathbf{k}}(\mathbf{r + R})  $, and a Bloch plane-wave $\mathrm{e}^{i \mathbf{k} . \mathbf{r}}$:

\begin{equation}
\Psi_{n,\mathbf{k}}(\mathbf{r}) = u_{n,\mathbf{k}}(\mathbf{r}).\mathrm{e}^{i \mathbf{k} . \mathbf{r}}.
\label{eqBlochBulk}
\end{equation}
The eigenstate $\Psi_{n,\mathbf{k}}(\mathbf{r}) $ varies continuously with the wave vector $\mathbf{k}$ and forms an energy band $\epsilon_{n,\mathbf{k}}$ identified by the band index $n$.

How does the band-structure picture change when the size of the system is reduced? Or, in other words, is the concept of energy bands also valid for nanoscopic systems? 
As we will show in the following, the finite size of a system has two important consequences: (i) wave vector $\mathbf{k}$ becomes discrete and (ii) wave functions are delocalized in momentum space \cite{HybertsenPRL1994}.

The first statement (i) directly follows from the construction of reciprocal space. For example, let us assume a 1-D mono-atomic chain of $N$ atoms with the lattice constant $a$. There are $N$ different wave vectors $\mathbf{k}$ separated by  
\begin{equation}
\Delta k^{(1)} = {2\pi \over a.N} = {2\pi \over L},
\label{eqDelta1}
\end{equation}
where $L=N.a$ is the length of the chain. Obviously, if $N$ becomes small, the separation between the wave vectors $\Delta k^{(1)}$ becomes larger. 

\begin{figure*}[!!!t]
\centering
\includegraphics[scale=0.5]{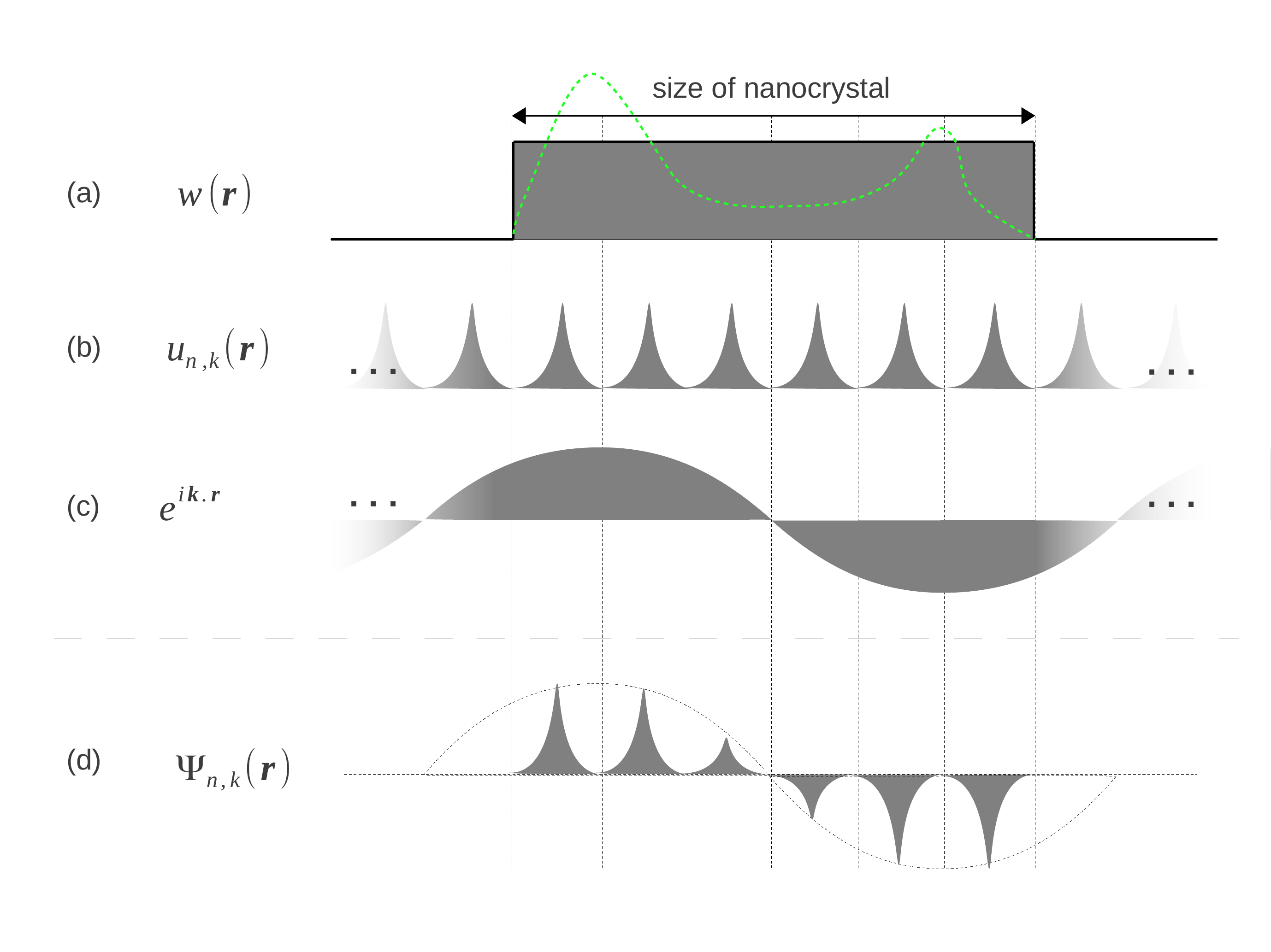}
\caption{ Schematic illustration of \refeq{eqBlochFinite}. (a) Window function constraining the wave function to a finite area of space. In general, it is the envelope of a MO and it can have a complicated shape (green dashed line). However, for simplicity, we expect the rectangular shape of \refeq{eqRect} (gray rectangle). (b) Mother function of an infinite system composed of a linear combination of atomic orbitals. (c) Infinite Bloch plane wave. (d) Final wave function of a nanocrystal as a product of (a),(b) and (c) constrained in a finite area of space.}
\label{Psi_Real}
\end{figure*}

Secondly, according to the Heisenberg uncertainty principle the crystal momentum becomes delocalized as follows: 
\begin{equation}
\Delta x \Delta p > h.
\label{eqHeisenberg}
\end{equation}
Substituting the crystal momentum $ p = \frac{h}{2 \pi } k $ and the size of the nanocrystal $\Delta x = L$ into the equation above, we obtain the relation for the delocalization of wave vector due to the finite size of the system

\begin{equation}
\Delta k^{(2)} > {2\pi \over L}.
\label{eqDelta2}
\end{equation}
The fact that the separation of discretized wave vectors $\Delta k^{(1)}$ and their delocalization $\Delta k^{(2)}$ are of the same order calls for a more rigorous discussion of the band structure of finite systems, which is provided in the following text. We will use natural units ($\hbar=1$), using $\mathbf{k}$ as a synonym for momentum.  
To simulate the finite size of systems (e.g.\ a nanocrystal) we introduce the so-called window function $ w(\mathbf{r}) $ which restricts the wave function $\Psi_{n,\mathbf{k}}
(\mathbf{r}) $ to the space occupied by the system. Then, \refeq{eqBlochBulk} is modified accordingly: 

\begin{equation}
\Psi_{n,\mathbf{k}}(\mathbf{r}) = w(\mathbf{r}) . u_{n,\mathbf{k}}(\mathbf{r}).\mathrm{e}^{i \mathbf{k} . \mathbf{r}}.
\label{eqBlochFinite}
\end{equation}
\reffig{Psi_Real} depicts the characteristics of the individual terms in \refeq{eqBlochFinite} in a 1-D case. In the simplest approximation, one can define the window function $w(\mathbf{r})$ as a~stepwise function

\begin{equation}
   w(x) = \left\{
     \begin{array}{lr}
       1 :  |\mathbf{r}| < L \\
       0 :  |\mathbf{r}| > L, \\
     \end{array}
   \right.
\label{eqRect}
\end{equation}
which vanishes outside the nanocrystal (see \reffig{Psi_Real}a). Then, the resulting real-space wave function $\Psi_{n,\mathbf{k}}(\mathbf{r})$ consists of the modulation of the mother function by the Bloch plane wave and the window function $ w(\mathbf{r})$ (see  \reffig{Psi_Real}d). In real systems, the window function can have a more complicated shape dictated by the electronic structure of a particular system (see green dashed line in \reffig{Psi_Real}a). 

\begin{figure*}[!!!t]
\centering
\includegraphics[scale=0.5]{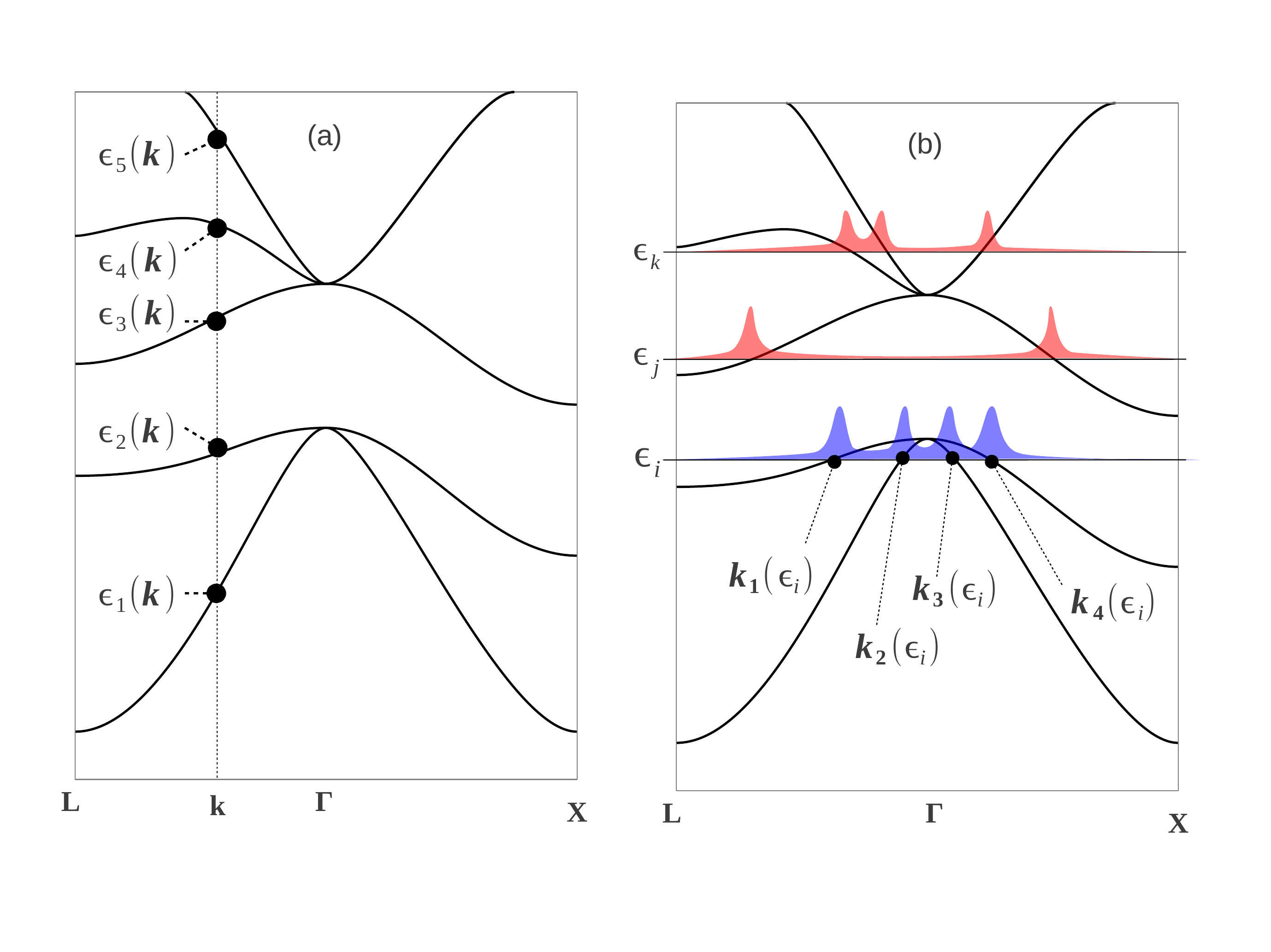}
\caption{ Schematic illustration of \refeq{eqPhi2Psi}. (a) Usual approach to obtaining the band structure of an infinite crystal using Bloch's theorem. For a given $\mathbf{k}$-vector, the energy spectrum of $\epsilon_{n,\mathbf{k}}$ is found. (b) Inverse approach in a finite system. For a given energy state $\epsilon_i$ of orbital $\phi_i$, the corresponding $\mathbf{k}$-vectors need to be found. The orbital is a linear combination of all Bloch-like wave functions $\Psi_{n,\mathbf{k}}(\mathbf{r})$ corresponding to the $n,\mathbf{k}$ in which the orbital energy  $\epsilon_i$ crosses the energy bands $\epsilon_n(\mathbf{k})$. }
\label{Linear_combination}
\end{figure*}

Let us consider a problem inverse to the computation of band structure from Bloch's theorem (\reffig{Linear_combination}a). We know the real-space wave function $\phi_i(\mathbf{r}) $ of $i$-th eigenstate (e.g. a~Kohn-Sham MO) obtained from an aperiodic DFT calculation of a nanoscopic system) and we would like to assign the corresponding crystal momentum $\mathbf{k}$. Each MO of energy $\epsilon_i$ can cross one or more bands $\epsilon_{n,\mathbf{k}}$ (\reffig{Linear_combination}b). Then, we can assume the MO $\phi_i(\mathbf{r}) $ to be a linear combination of $A$ bands $\Psi_{n,\mathbf{k}}(\mathbf{r})$:

\begin{equation}
\phi_i(\mathbf{r}) = \sum\limits_{a=0}^A d_a  \Psi_{n_a,\mathbf{k}_a}(\mathbf{r}),
\label{eqPhi2Psi}
\end{equation}
where, for a particular $\phi_i(\mathbf{r})$, neither the indexes $n_a,\mathbf{k}_a$ nor the coefficients $ d_a $ are explicitly known.

One option how to extract $\mathbf{k}$-vectors of $\phi_i(\mathbf{r})$ lies in the determination of the corresponding $\Psi_{n_a,\mathbf{k}_a}(\mathbf{r})$ and $u_{n_a,\mathbf{k}_a}(\mathbf{r})$ and a consecutive extraction of the Bloch factor $\mathrm{e}^{i \mathbf{k} . \mathbf{r}}$ from \refeq{eqBlochFinite}. This is possible, for example, by projecting $\phi_i(\mathbf{r})$ on $\Psi^{BULK}_{n,\mathbf{k}}(\mathbf{r})$ obtained from bulk calculation \cite{ValentinPhonon}.
Nevertheless, wave functions $\Psi_{n,\mathbf{k}}(\mathbf{r})$ in a finite system may differ considerably from the bulk one $\Psi^{BULK}_{n,\mathbf{k}}(\mathbf{r})$ (e.g.\ if $w(\mathbf{r})$ is more complicated) and a large number of states have to be considered for a large nanocrystal. These complications hamper a~robust implementation of this method to a computational code.

Here we opt for another method, which transforms MO $\phi_i(\mathbf{r})$ from  real into reciprocal space using Fourier transform. A similar approach was successfully applied to analyze angular resolved photoemission (ARPES) spectra of oligomeric organic molecules \cite{PuschnigScience09}. In our scheme, selected $\phi_i(\mathbf{r})$ expanded on a real space grid is projected to  momentum space either using discrete 3-D Fast Fourier Transform (FFT) or by the projection onto a set  of plane waves $\mathrm{e}^{i \mathbf{k'} . \mathbf{r}}$:

\begin{equation}
\tilde{\phi}_i(\mathbf{k'}) = \langle \phi_i(\mathbf{r}) | \mathrm{e}^{i \mathbf{k'} . \mathbf{r}} \rangle,
\label{eqCProject}    
\end{equation}
where $\mathbf{k'}$ is an arbitrary wave vector.

\begin{figure*}[!!!t]
\centering
\includegraphics[scale=0.5]{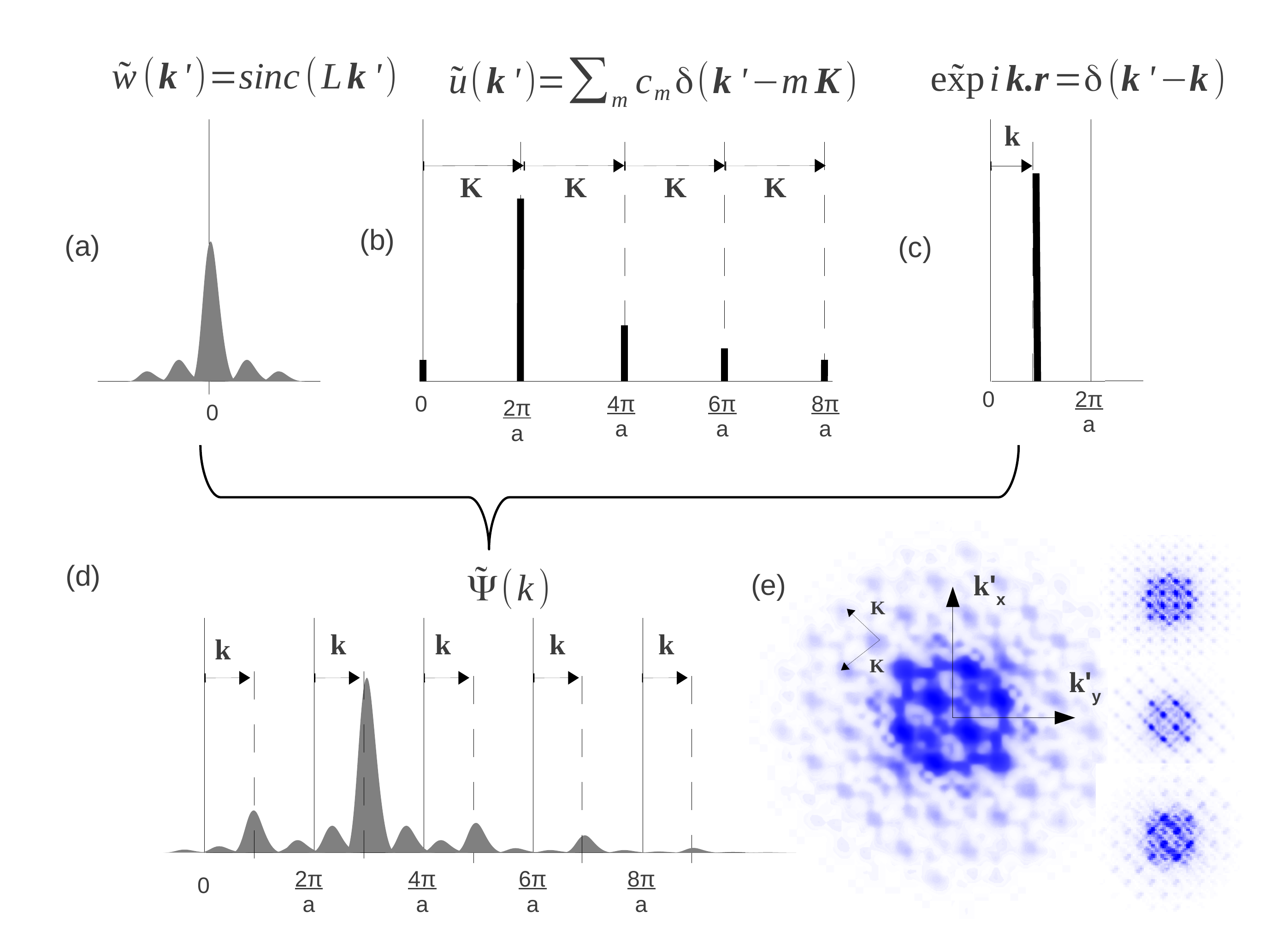}
\caption{ Schematic illustration of convolution of momentum space representations of the three wave function components. (a) Rectangular window function transformed to $\mathrm{sinc}(L \mathbf{k'})$. (b) Bulk-like mother function transformed to a series of $\delta$-functions in the centers of Brillouin zones. (c) Bloch wave transformed to a $\delta$-function inside the first Brillouin zone shifted out of the center. (d) Momentum space representation of the wave function created by the convolution of (a),(b) and (c) in a 1-D case. (e) Illustration of a more-D case: a 2D view of momentum space density of MOs in a silicon nanocrystal. Note the side artifacts spaced by multiples of $ K $ representing the higher Fourier components $c_m$ in the expansion of mother function.}
\label{k-convolution}
\end{figure*}

For practical use, we plot the so-called momentum density $\tilde{\rho}_{i}(\mathbf{k'})$, which is real, instead of the complex Fourier transform $\tilde{\phi}_i(\mathbf{k'})$. The momentum density $\tilde{\rho}_{i}(\mathbf{k'})$ can be written, using \refeq{eqCProject}, as follows: 

\begin{equation}
\tilde{\rho}_{i}(\mathbf{k'}) = | \tilde{\phi}_i(\mathbf{k'}) |^2.   
\label{eqCdens}
\end{equation}
It also contains the information about the delocalization of the MO in momentum space and it can be plotted as a~function of wave vector along a~selected line in momentum space (see \reffig{k-convolution}d).

The first approach (FFT) provides a picture of the 3-D structure and symmetries of the particular state in reciprocal space (see \reffig{k-convolution}e). 
Nevertheless, its resolution is limited by the size of the real-space grid on which MOs are expanded. The second approach (the projection on set of plane waves) is more suitable to stick with the traditional 1-D band 
structure representation plotted along the lines connecting the high-symmetry points in $\mathbf{k}$-space. In this case, we let $\mathbf{k'}$ sample the given high-symmetry line in 
$\mathbf{k}$-space with much higher resolution. Although this approach is straightforward, a~rigorous analysis of the structure of the resulting momentum space distribution is 
required. For simplicity, let us assume a one-to-one correspondence between a particular MO and a band wave function $ \phi_i(\mathbf{r}) = \Psi_{n,\mathbf{k}}(\mathbf{r})$. Note that the generalization of the following discussion

if  $\phi_i(\mathbf{r})$ is  
a linear combination of several $\Psi_{n_a,\mathbf{k}_a}(\mathbf{r})$ (\refeq{eqPhi2Psi}) 
is straightforward due to the linearity of Fourier transform.
Fourier transform of a wave function $\Psi_{n,\mathbf{k}}(\mathbf{r})$ given by \refeq{eqBlochFinite} can be expressed as a convolution of the three terms:

\begin{equation}
\tilde{\Psi}_{n,\mathbf{k}}(\mathbf{k'}) = \tilde{w}(\mathbf{k'}) \star  \tilde{u}(\mathbf{k'}) \star \delta(\mathbf{k' - k}).
\label{eqBlochConv}
\end{equation}
It is well worth analyzing in detail the process of convolution and the character of each term in \refeq{eqBlochConv}. To make our discussion more illustrative, we restrict ourselves to 1-D case. 
\reffig{k-convolution} represents schematically the process of convolution and the character of each term separately. 

We start from the trivial third term. $ \delta (\mathbf{k'-k}) $ is a Fourier transform of Bloch wave $ \mathrm{e}^{i\mathbf{k.r}} $, where $\mathbf{k}$ is from the first Brillouin zone (see \reffig{k-convolution}c) and attains some of the discrete values separated by $ \Delta k^{(1)} = 2\pi/L $ as postulated in \refeq{eqDelta1}.

The first term in $\tilde{w}(\mathbf{k'})$ in \refeq{eqBlochConv} is a Fourier image of the window function $w(\mathbf{r})$. This term causes the delocalization of wave function in $\mathbf{k}$-space. Considering the  window function $w(\mathbf{r})$ as the step function of length $L$ (see \refeq{eqRect}), its Fourier transform equals 
$ \tilde{w}(\mathbf{k'}) = \mathrm{sinc}(L \mathbf{k'})$. Using \refeq{eqDelta2}, it can be expressed also as $\mathrm{sinc}( 2\pi \mathbf{k'} / \Delta k^{(2)})$, which brings us back to the Heisenberg principle.

The square $|\tilde{w}(\mathbf{k'})|^2$ introduces to the  momentum space density 
$ \tilde{\rho_i}(\mathbf{k'})$ a blur with a Lorentzian envelope (see \refeq{k-convolution}a) of slow asymptotic decay $1/ (L \mathbf{k'})^2$. This has an important implication 
for the momentum selection rules for the transitions between states in finite systems: two wave functions centered around different $\mathbf{k}$ can have non-negligible overlap 
in $\mathbf{k}$-space even if their separation in $\mathbf{k}$ is fairly large. 

We would like to stress that the decay in the momentum space projection depends strongly on the actual shape of $w(\mathbf{r})$. Therefore, this approach is 
superior to a simple estimation of momentum uncertainty based on the Heisenberg principle, which can provide only the width of the peak around $\mathbf{k}$, but says nothing 
about its decay. 

Finally, the second term in \refeq{eqBlochConv}  $\tilde{u}(\mathbf{k'})$ corresponds to the Fourier transform of the mother function $u_{n,\mathbf{k}}(\mathbf{r})$, which can be expanded in a discrete Fourier series of waves using its periodicity: 

\begin{equation}
u_{n,\mathbf{k}}(\mathbf{r}) = \sum\limits_{m=0}^\infty c_m  \mathrm{e}^{i( m.\mathbf{K}) . \mathbf{r}},
\label{eqUexpansion_real}
\end{equation}
where $ \mathbf{K} $ denotes the reciprocal lattice vector and $m$ is an integer index addressing different reciprocal unit cells. This expression is transformed to reciprocal space as follows: 

\begin{equation}
\tilde{u}_{n,\mathbf{k}}(\mathbf{k'}) = \langle \sum\limits_{m=0}^\infty c_m  \mathrm{e}^{i( m.\mathbf{K}) . \mathbf{r}} | \mathrm{e}^{i \mathbf{k'} . \mathbf{r}} \rangle = \sum\limits_{m=0}^\infty c_m  \delta(\mathbf{k'}-m.\mathbf{K}),
\label{eqUexpansion}
\end{equation}
where $\delta(\mathbf{k'}-m.\mathbf{K})$ is situated in the center of the $m$-th reciprocal unit cell (i.e.\ not in the first Brillouin zone). 

\begin{figure*}[!!!t]
\centering
\includegraphics[scale=0.5]{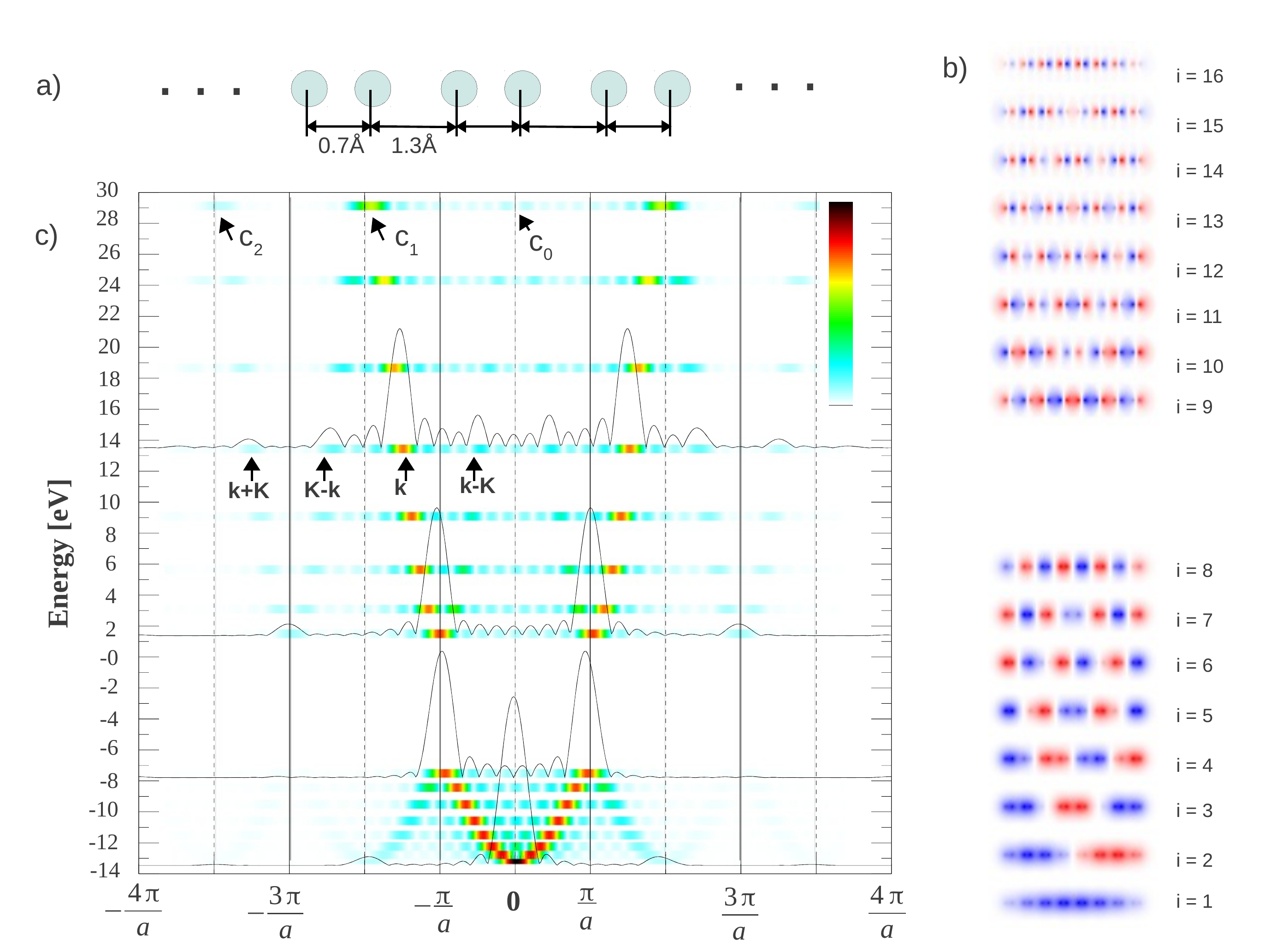}
\caption{ 1-D cross-section of momentum-space projected MOs of a chain of 8 hydrogen molecules. (a) Atomic structure of molecular hydrogen chain with bond lengths of 0.7 \AA electronically coupled through the intermolecular distance of 1.3 \AA. (b) Real space projection of MOs $ \phi_i(\mathbf{r}) $ of the system. First 8 orbitals are bonding and occupied, the remaining 8 are anti-bonding and unoccupied. Positive sign of wave function is colored in blue, the negative sign in red. (c) Band structure of the finite chain as assembled from momentum densities  $ \rho_i(\mathbf{k'}) $ plotted at particular orbital energy $\epsilon_i$ by color scale. Parabolic band dispersion can be clearly seen following the maxima of densities (yellow-red-black). For some of the orbitals $i=1,8,9,13$, the actual shape of function $ \rho_i(\mathbf{k'}) $ is also plotted to illustrate the correspondence to \reffig{k-convolution}. The main peaks corresponding to the Fourier components in the expansion 
of $u_{n,\mathbf{k}}(\mathbf{k'})$ are denoted by arrows. }
\label{H-chain}
\end{figure*}

Typically, $ u_{n,\mathbf{k}}(\mathbf{r}) $ is relatively smooth in the unit cell and so the Fourier expansion coefficients $ c_m $ decay with increasing 
frequency (\reffig{k-convolution}b). In reality, just one expansion coefficient is dominant. The dominant Fourier component is determined by the nodal structure of 
$ u_{n,\mathbf{k}} (\mathbf{r}) $, where nodes are introduced either by anti-bonding character of the  mother function or by the contribution of the higher angular momentum atomic 
orbitals (such as a $p$-orbital). If $ u_{n,\mathbf{k}}(\mathbf{r}) $ has no node inside a unit cell (for example bonding state of two $s$-orbitals), the first coefficient $c_0 $ 
situated in the center of the first Brillouin zone ($ \mathbf{k'}=0 $) is the dominant term. Similarly, if $ u_{n,\mathbf{k}}(\mathbf{r}) $ has $m$ nodes,  the $m$-th Fourier 
expansion coefficient $c_m$ dominates. This means that most of the momentum space density is situated in the $m$-th reciprocal unit cell. Consequently, the 
convolution of  $ \tilde{u}_{n,\mathbf{k}}(\mathbf{k'})$ with the Bloch term $ \delta (\mathbf{k'-k}) $ forms a new band located in the $m$-th reciprocal unit cell 
(see the anti-bonding band in \reffig{H-chain}c). Thus, our approach provides the reconstruction of the so-called \textit{unfolded} band structure, where each band is situated in a different reciprocal unit cell. Note also that in a 3-D case the index $m$ is a vector, and nodes in each dimension should be considered independently.

Now, we will illustrate the projection method on a~simple case. Let us analyze the electronic structure of a finite 1-D chain consisting of 8 coupled hydrogen molecules 
(see \reffig{H-chain}), where the unit cell consists of a~single hydrogen molecule. Upon projecting real space MOs (\reffig{H-chain}b) into $\mathbf{k}$-space two distinct bands isolated 
by a band gap appear (see \reffig{H-chain}c). The lower band ($m=0$) composed of bonding orbitals with no node inside the unit cell has the dominant Fourier components 
located in the first Brillouin zone. The higher anti-bonding band ($m=1$) has the dominant Fourier components located in the second reciprocal unit cell. 

Apart from these dominant Fourier components, which characterize the band structure, there are also smaller satellite peaks representing the other components $c_m$ in the Fourier expansion of $\tilde{u}_{n,\mathbf{k}}(\mathbf{k'})$. In the special case, when Bloch $\mathbf{k}=0$, the peaks are located directly in centers of $m$-th reciprocal unit cell and correspond directly to components $c_m$ in Fourier expansion of $\tilde{u}_{n,\mathbf{k}}(\mathbf{k'})$ (\refeq{eqUexpansion}). This is illustrated for MO $i=16$ by peaks denoted $c_0, c_1, c_2$ in \reffig{H-chain}c. In general case ($\mathbf{k} \neq 0$ ) these peaks are split and shifted by $\pm \mathbf{k}$ as denoted by $\mathbf{K}+\mathbf{k}$ and $\mathbf{K}-\mathbf{k}$ in \reffig{H-chain}c.

\begin{figure*}[!!!t]
\centering
\includegraphics[scale=0.5]{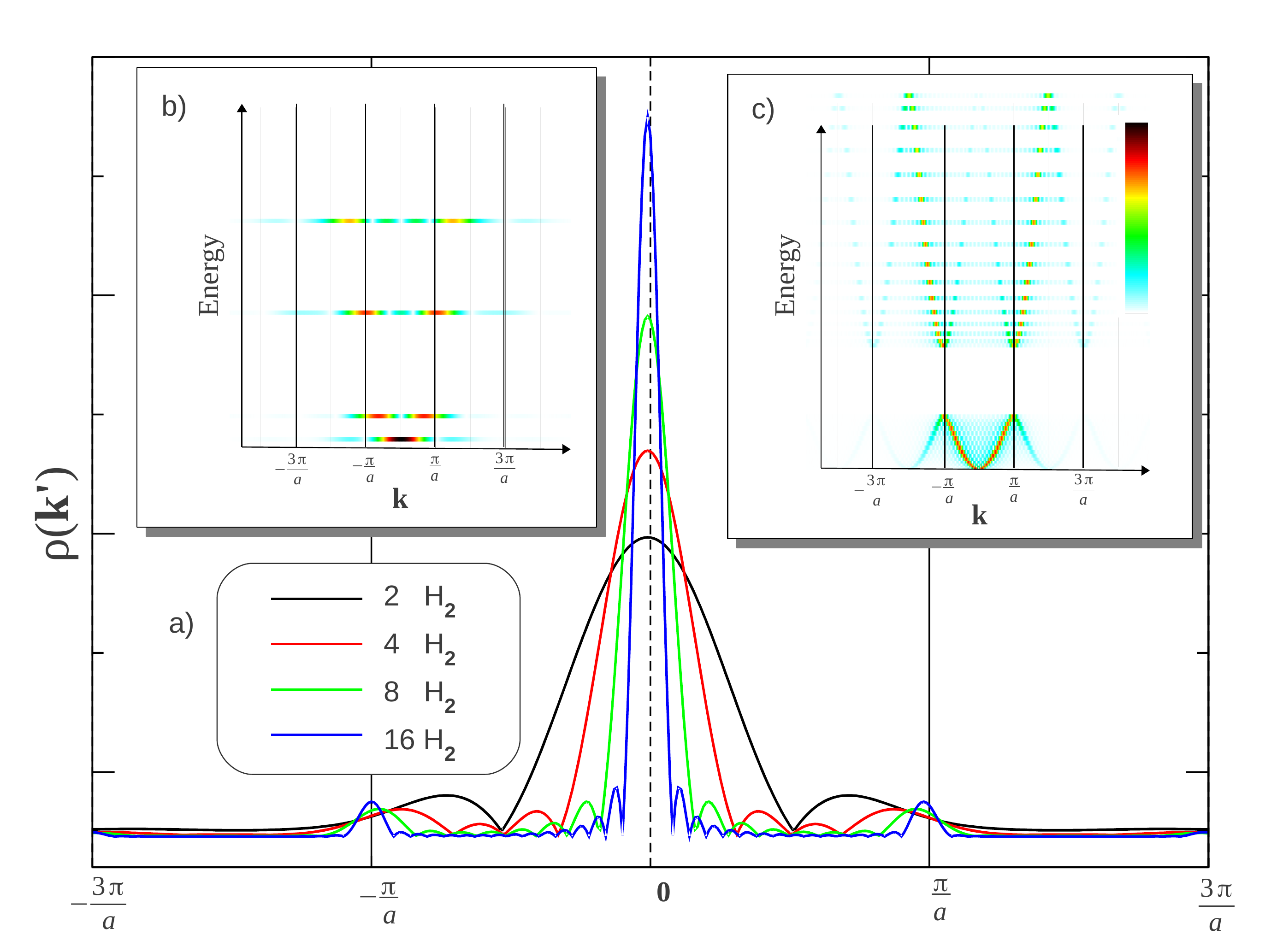}
\caption{ Dependence of the delocalization of momentum density  $\tilde{\rho}_{i}(\mathbf{k'})$ of MOs in a chain of hydrogen molecules on the chain length. (a) Momentum space delocalization of the lowest MO ($i=1$) in a chain composed of 2 (black), 4 (red), 8 (green) and 16 (blue) hydrogen molecules. (b) Band structure of a  chain composed of 2 hydrogen molecules. (c) Band structure of a chain composed of 16 hydrogen molecules. }
\label{H-chain-delocalization}
\end{figure*}

To illustrate the delocalization of the momentum vector $\mathbf{k}$ due to the finite size of the system, represented by $w(\mathbf{r})$, we analyzed the lowest band ( $i=1$ )
of several different 1-D hydrogen chains with lengths from 2 to 16 molecules.
In \reffig{H-chain-delocalization}a the peak width of the depicted momentum density decreases proportionally to the number of molecules (unit cells) in the chain according to \refeq{eqDelta2}. In this particular case, the window function $w(\mathbf{r})$ is very close to a rectangular step function. Hence, the shape of the momentum space density  
$\tilde{\rho}(\mathbf{k'})$ is very similar to $|\mathrm{sinc}( L \mathbf{k'})|^2$. The frequency of this $\mathrm{sinc}$-like function can be also deduced from the number of nodes per reciprocal unit cell, which is proportional to the increasing chain length $L$.

Let us summarize the main conclusions of this chapter. We discussed two main implications of the finite size of a nanocrystal on its band structure: (i) the discretization and (ii) the delocalization of wave vector $\mathbf{k}$. We introduced a robust method of projection of MOs to $\mathbf{k}$-space, which provides band structure of a finite system, and discussed in detail the analysis of the resulting momentum densities $\tilde{\rho}_{i}(\mathbf{k'})$. We demonstrated that the important characteristics (i.e.\ Bloch $\mathbf{k}$-vector, its delocalization and the discrimination of independent bands) can be extracted from the resulting momentum densities, despite the convolution in \refeq{eqBlochConv}.

\subsection{ Band structure of bulk silicon and nanocrystal }

\begin{figure*}[!!!t]
\centering
\includegraphics[scale=0.5]{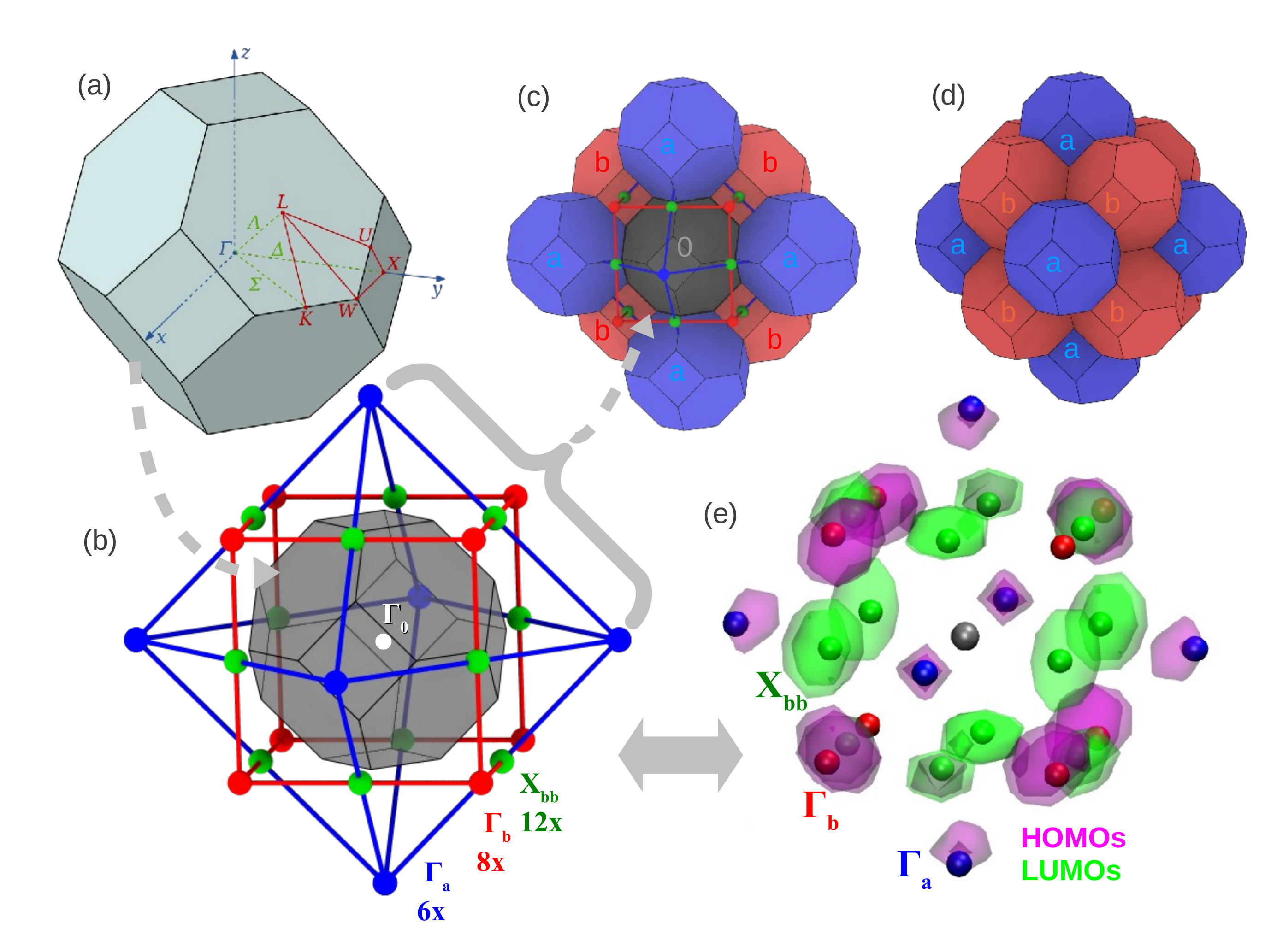}
\caption{ 3D structure of silicon reciprocal lattice. 
(a) Structure of the first Brillouin zone of silicon in the  shape of a truncated octahedron with high symmetry directions shown. 
(b) Positions of points $\Gamma_a$ (blue), $\Gamma_b$ (red) and $X_{bb}$ (green) around the  first Brillouin zone.  
(c) The points $\Gamma_a$, $\Gamma_b$ and $X_{bb}$ in the context of the sub-lattices {\bf a}(blue) and {\bf b} (red). 
(d) Reciprocal unit cells adjacent to the first Brillouin zone, distinguished to the sub-lattice {\bf a} (blue) and {\bf b} (red).
(e) 3D view of the localization of maximal $\mathbf k$-space projected density of HOMOs (green) and LUMOs (purple) for a Si$_{68}$-H \nc.  Points $\Gamma_a$, $\Gamma_b$ and $X_{bb}$ marked by spheres to highlight the correspondence.}
\label{honeycomb}
\end{figure*}

Before we advance from the 1-D hydrogen chain to the band structure of a silicon \nc, it is worth discussing in detail the reciprocal space structure of bulk silicon. While in solid state physics it is common to describe band structure in the compact picture (or reduced zone scheme), where all bands are folded into the first Brillouin zone, for our purposes it is more suitable to use the so-called unfolded picture. In this approach, bands located in higher reciprocal unit cells are considered independently. 

Silicon crystallizes in a  diamond lattice consisting of two interpenetrating face centered cubic (FCC) Bravais lattices. In real space, the Wigner-Seitz cells of FCC build up a rhombic icosahedral honeycomb, which converts to a truncated octahedral honeycomb in reciprocal space. As shown in \reffig{honeycomb}d, the reciprocal lattice is composed of two cubic sub-lattices {\bf a} (blue) and {\bf b} (red) with a truncated octahedral shape. The sub-lattice {\bf a} is centered in $\Gamma_0$ of the first Brillouin zone and the sub-lattice {\bf b} is shifted by a~vector (1,1,1) with respect to the sub-lattice {\bf a}.  Thus, the sub-lattice {\bf b} is situated in the cube vertexes and the sub-lattice {\bf a} in the cube centers of body centered cubic lattice (BCC), as shown in \reffig{honeycomb}b.

The unit cell located at $\Gamma_0$  is the first Brillouin zone. There are in total 14 reciprocal unit cells adjacent to the first Brillouin zone. 
Six of them belong to the sub-lattice {\bf a}, being centered at the  points $\Gamma_a$ with coordinates $\forall$ ($\pm 2$,0,0). Another 8 unit cells belongs to the sub-lattice {\bf b} centered at $\Gamma_b$ with coordinates $\forall$ ($\pm 1$,$\pm 1$,$\pm 1$). Here, $\forall$ means all permutations of axes and signs (e.g.\ ($\pm 2$,0,0), (0,$\pm 2$,0) and (0,0,$\pm 2$) ).  
 
All these 14 reciprocal unit cells are of crucial importance for us, because both the valence and conduction bands are situated there (in \textit{unfolded} picture). 
In particular, the valence band maximum is located in the points $\Gamma_a$ and $\Gamma_b$. The {\it absolute} conduction band minimum in bulk silicon is located near the points $X_{bb}$ of coordinates $\forall$ ($\pm 1$,$\pm 1$,0), which are halfway between two $\Gamma_b$. The positions of the points $\Gamma_a$, $\Gamma_b$ and $X_{bb}$ are schematically 
depicted in \reffig{honeycomb}b. The localization of maximal $\mathbf k$-space projected density obtained from HOMOs (purble) and LUMOs (green) for a Si$_{68}$-H \nc is shown in \reffig{honeycomb}e. From this figure, it is evident that HOMOs are localized at $\Gamma_a$, $\Gamma_b$ and LUMOs near the $X_{bb}$ points. 
To make the context more clear, in \reffig{honeycomb}c we also depict the positions of the important points $\Gamma_a$, $\Gamma_b$ and $X_{bb}$ in the context of the sub-lattices {\bf a} and {\bf b}. Because of convolution with Fourier expansion of mother function ( \refeq{eqUexpansion} or more illustrative \reffig{k-convolution}b,d ) both the valence and the conduction bands are situated outside of the first Brillouin zone in \textit{unfolded} picture. Therefore, the first Brillouin zone does not play any significant role in optical transitions and it is depicted in gray in \reffig{honeycomb}b,c.

In the case of bulk silicon, all these reciprocal unit cells around $\Gamma_a$ and $\Gamma_b$ are equivalent due to lattice symmetry. Therefore also the frontier orbitals (HOMOs and LUMOs), which are composed of Bloch states located in $\Gamma_a$, $\Gamma_b$ and $X_{bb}$, are energetically degenerate. More precisely, HOMOs (resp. LUMOs) are arbitrary linear combinations of Bloch states from different $\Gamma_a$, $\Gamma_b$ (resp. $X_{bb}$) as was discussed in previous chapter ( \refeq{eqPhi2Psi} and \reffig{Linear_combination} ). 

However, this is no longer true in \nc s, where the symmetry is broken due to different lengths in each direction or due to anisotropic perturbations (e.g.\ mechanical strain) induced by surface passivation. This is another reason why it is necessary to describe the band structure of a nanocrystal in the \textit{unfolded} picture with all reciprocal unit cells independent.

In other words, the anisotropy of \nc s removes the degeneracy of MOs. We can observe this in our model \nc s even though the shape of cores (Si$_{68}$, Si$_{232}$, Si$_{538}$) respects the cubic symmetry of silicon lattice. In the case of --H passivation, the degeneracy is almost preserved, because no anisotropic strain or electrostatic field is induced. The LUMO state is 12-fold degenerate, representing the  12 points $X_{bb}$. 

\begin{figure*}[!!!t]
\centering
\includegraphics[scale=0.5]{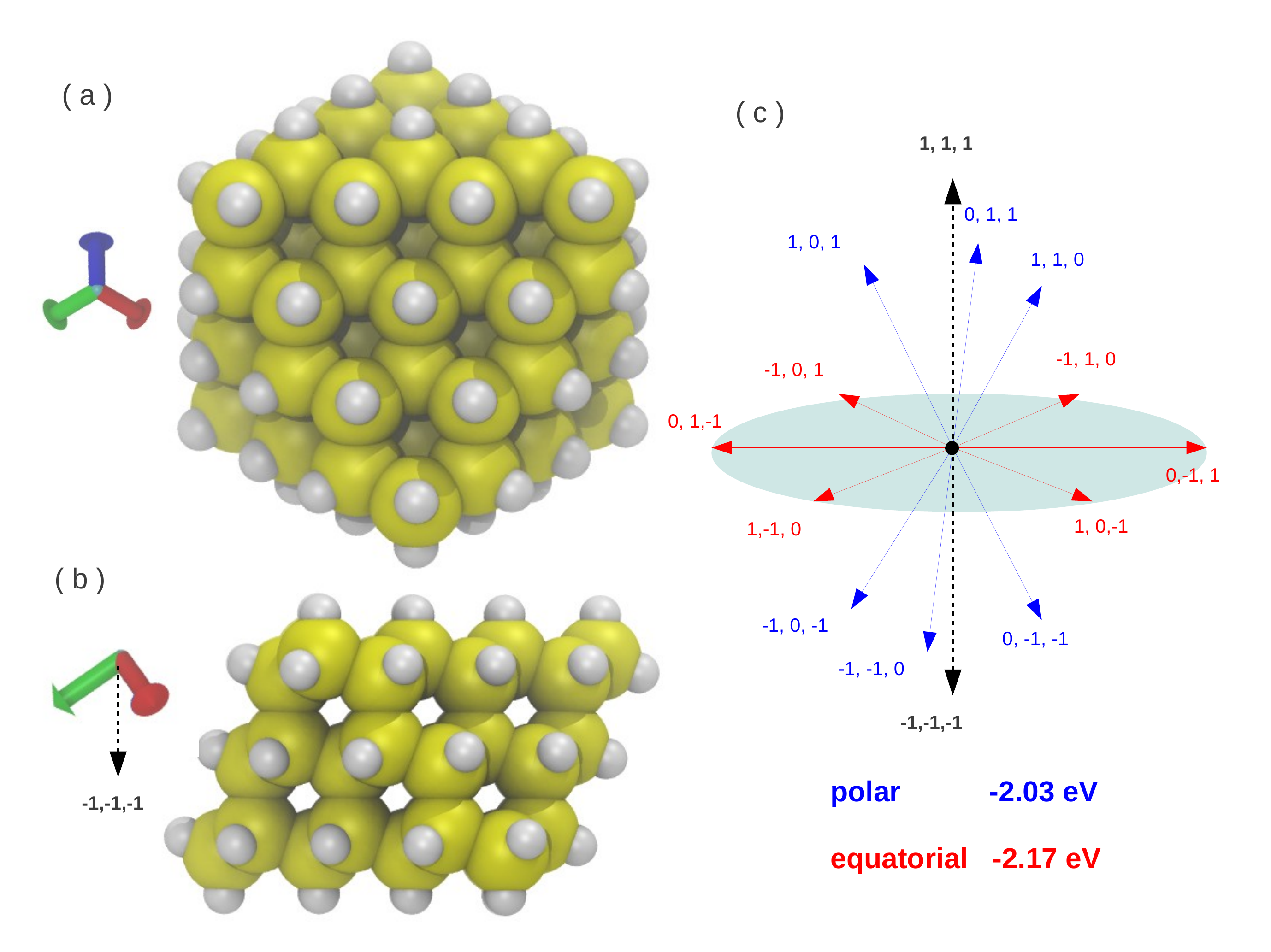}
\caption{ Degeneracy loss due to the broken symmetry of a Si$_{68}$-H nanocrystal. 
(a) Top view of a Si$_{68}$-H nanocrystal along the 111-direction. 
(b) Side view of a Si$_{68}$-H nanocrystal perpendicularly to the 111-direction. 
(c) Two classes of energy split MOs of Si$_{68}$-H due to anisotropic confinement. There are 6 degenerate states in the directions closer to 111-direction (polar, blue) with the energy of 2.17 eV, and other 6 degenerate states more perpendicular to the 111-direction with the energy of $-2.03$ eV (equatorial, red). Note that states $\mathbf{k}$ and $-\mathbf{k}$ are always degenerate due to time reversal symmetry and the realness of wave function in a finite system. }
\label{Symmetry_Breaking}
\end{figure*}

However, due to the shape of a Si$_{68}$-H nanocrystal---slightly shorter in the (111) direction  (\reffig{Symmetry_Breaking}a), MOs are split into two classes of slightly different energies. Electronic states with $\mathbf k$-vector oriented closer to (111) have energy of 2.17~eV, while states with $\mathbf k$-vector oriented more perpendicularly to the (111) direction have energy of $-2.03$~eV (see \reffig{Symmetry_Breaking}c). 

Finally, we make a comment on the importance of MO symmetry for optical transitions. Since MOs are real functions, their Fourier transforms are symmetric for $\mathbf{k}$ and $-\mathbf{k}$. Consequently, we can identify two distinct classes of MOs according to the character of their symmetry: (a) if a MO is an even function in real space, its $\mathbf k$-space representation becomes real (cosine part); (b) if a MO is an odd function, the $\mathbf k$-space representation is purely imaginary (sine part). 
According to selection rules, optical transitions are strongest between two states localized around the same point in $\mathbf k$-space with complementary symmetry character, i.e.\ sine to cosine or vice versa.

\section{ Results}
In this section, we will employ the method introduced in the previous section to analyze the band structure of Si \nc s as a~function of their diameter and surface passivation of Si \nc s of nanometer size. Namely, we will analyze an~impact of surface passivation including both polar (--OH) and non-polar (--H,--CH$_3$) groups on their atomic and electronic structure.

\begin{figure*}[!!!t]
\centering
\includegraphics[scale=0.5]{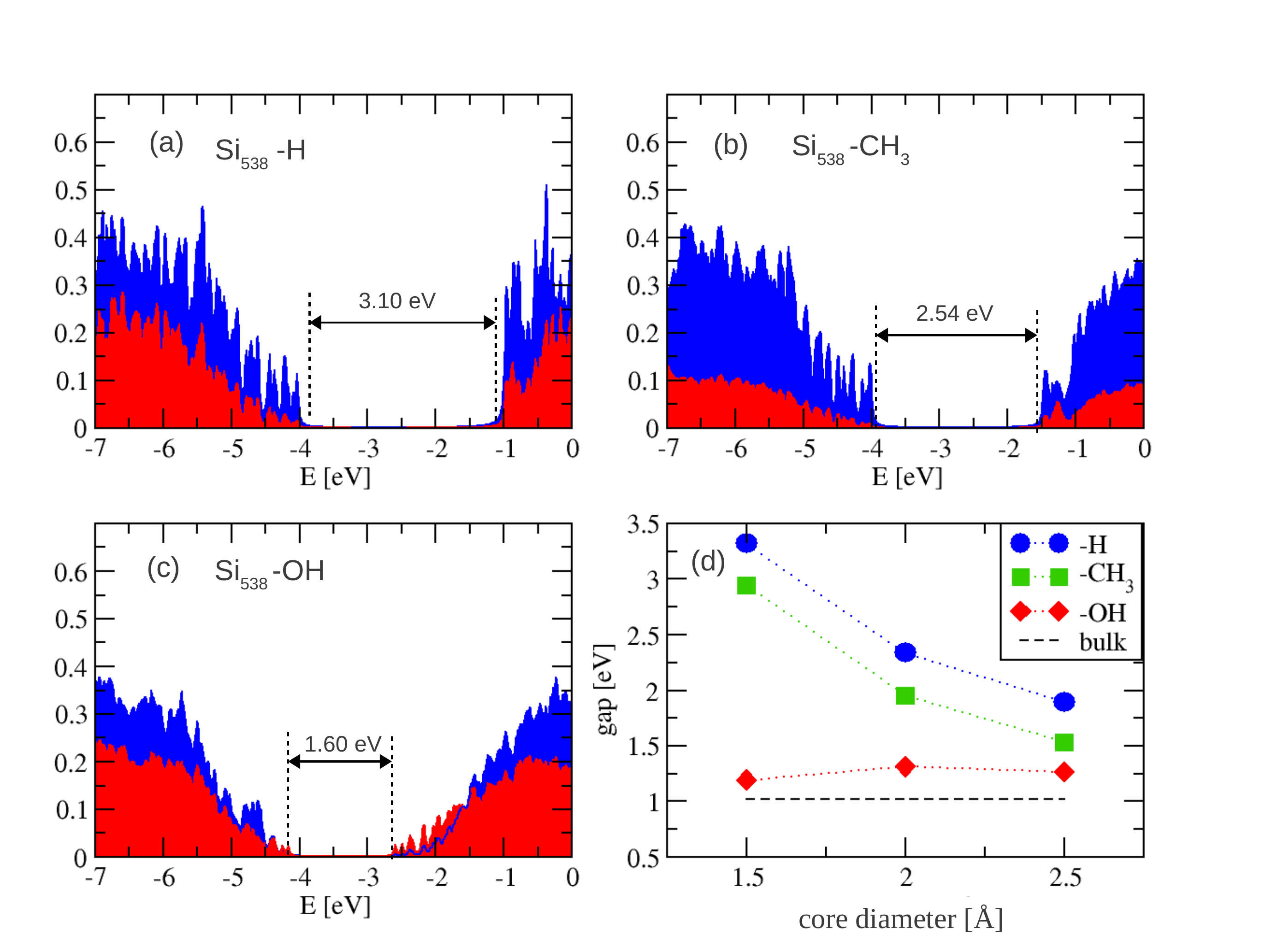}
\caption{ Projected density of states around gap split into the contributions of surface (red) and core atoms (blue), for the non-polar passivation (a) Si$_{538}$-H and (b) Si$_{538}$-CH$_3$; frontier orbitals are localized mostly inside the core. However, for an oxidized \nc\ (c) Si$_{538}$-OH, these are mostly surface states. (d) shows variation of energy HOMO-LUMO gap with core diameter for all three types of passivation }
\label{PDOS_gap}
\end{figure*}

\begin{figure*}[!!!t]
\centering
\includegraphics[scale=0.5]{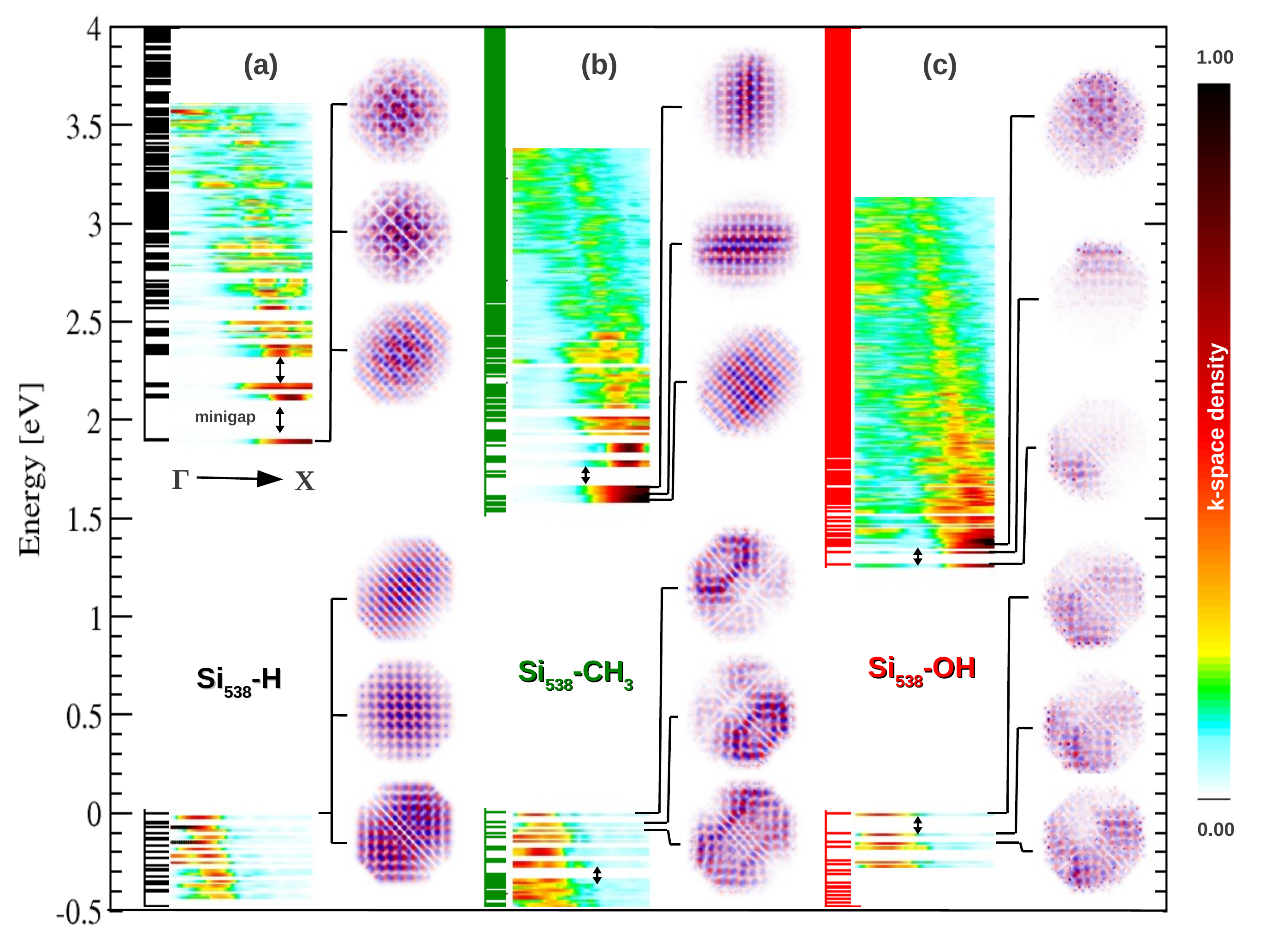}
\caption{ Spectrum of electronic states (the black, green and red bars) with their $\mathbf k$-space projection from $\Gamma$ to $X$ (color scale legend is on the right) for (a) Si$_{538}$-H, (b) Si$_{538}$-CH$_3$ and (c) Si$_{538}$-OH.  The real space projections of selected frontier orbitals are plotted as well in order to show their overall shape, symmetry and localization.  Note: colorscale was limited to the same maximal value in order to enhance the contrast and to compare the absolute $\mathbf k$-space densities between different passivations. }
\label{k_and_real}
\end{figure*}

\begin{figure*}[!!!t]
\centering
\includegraphics[scale=0.5]{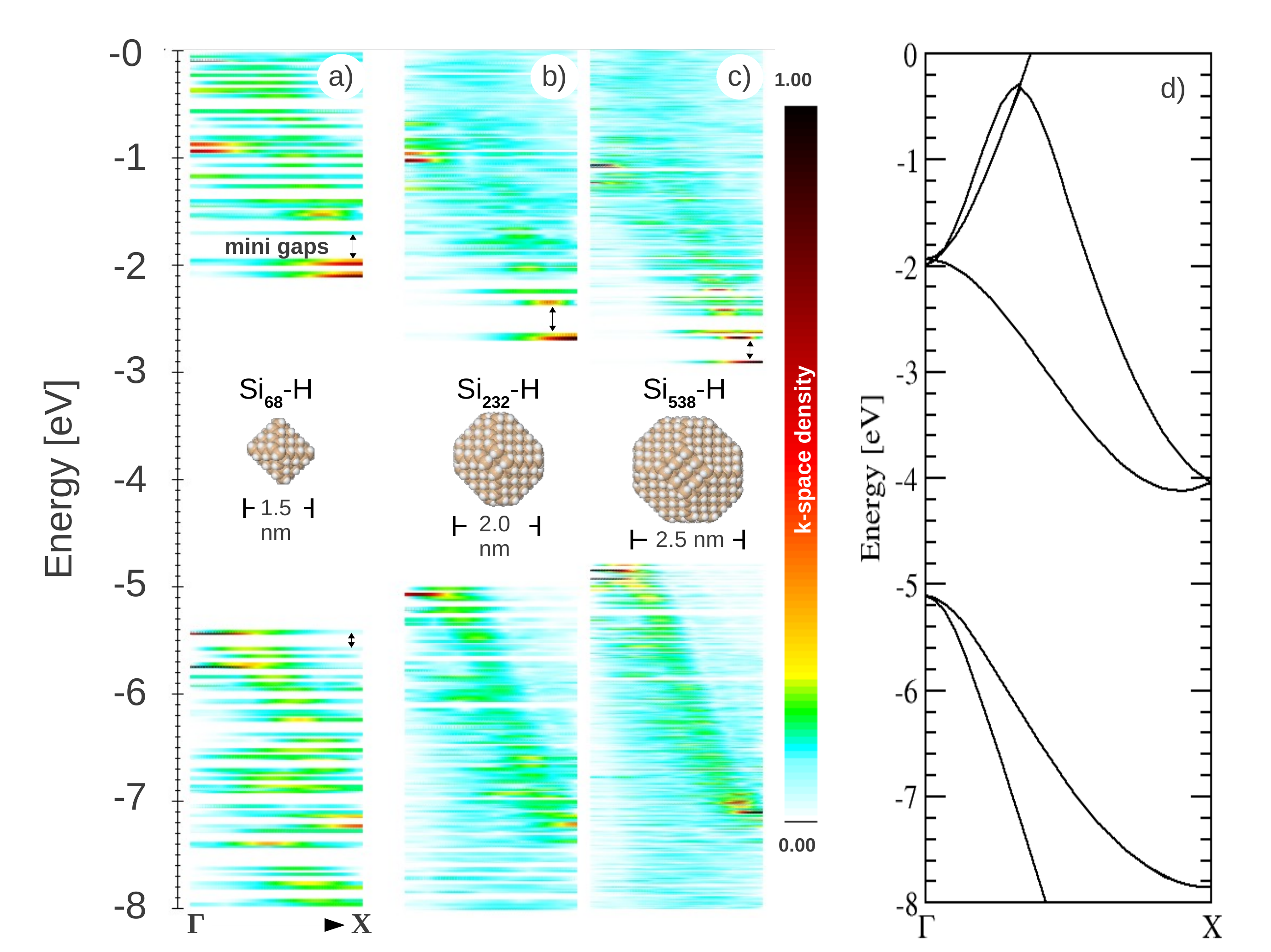}
\caption{ Comparisons of 1-D cross-sections of projected $\mathbf k$-space density in --H passivated \nc s of different sizes for (a) Si$_{68}$-H, (b) Si$_{232}$-H and (c) Si$_{538}$-H. Gradual emergence of energy bands can be seen with increasing \nc size. 
 (d) Band structure of bulk Si for comparison. Note: color scale was normalized by maximal density independently for each \nc\ size. }
\label{k-lines-size}
\end{figure*}

\begin{figure*}[!!!t]
\centering
\includegraphics[scale=0.5]{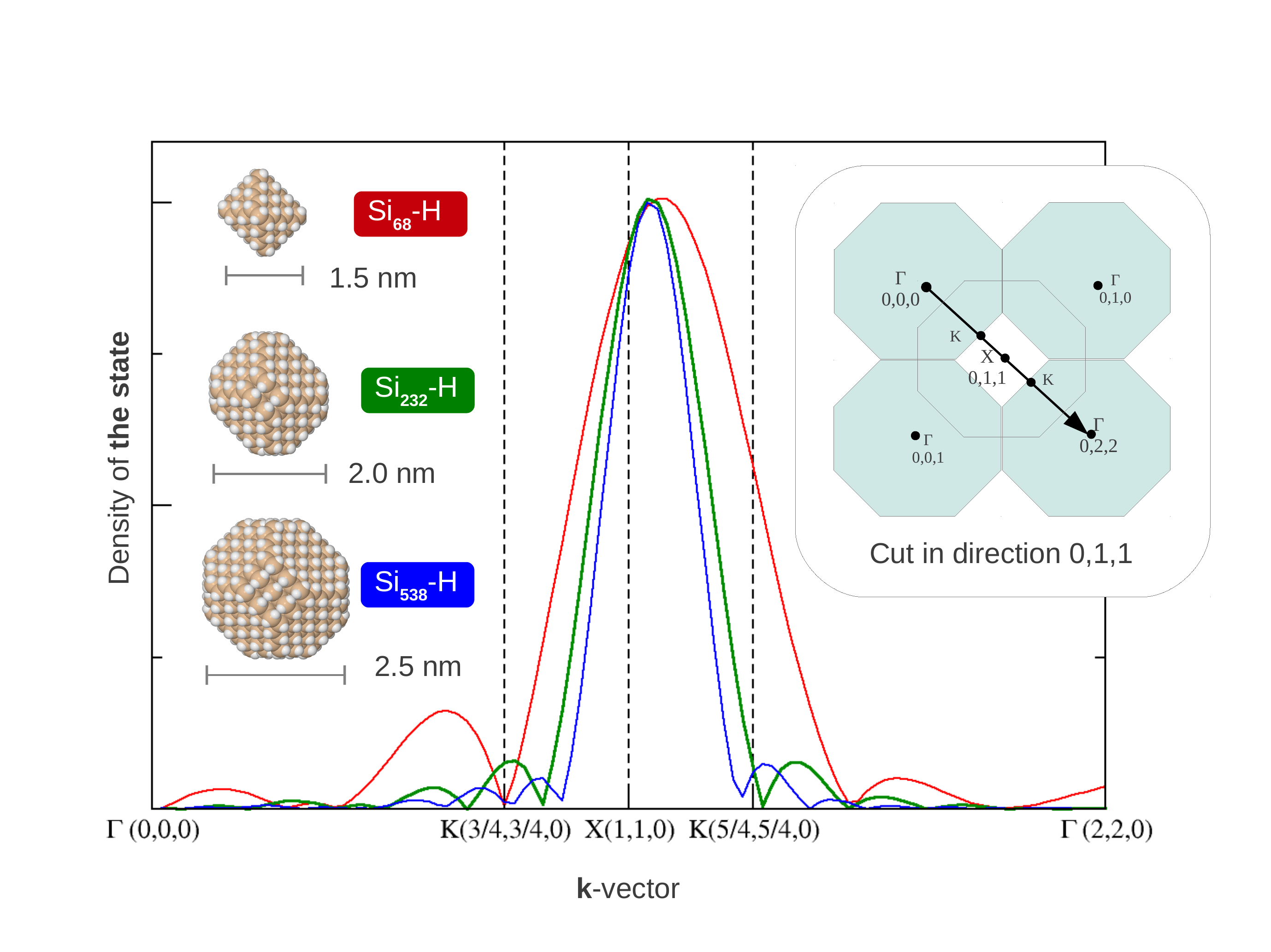}
\caption{ Delocalization of $\mathbf k$-space projection of LUMO in --H passivated \nc s depending on the nanocrystal diameter. The $\mathrm{sinc}$-like shape can be clearly seen, which means an approximately rectangular envelope of the MO in real space.  }
\label{k-delocalization_by_size}
\end{figure*}

\begin{figure*}[!!!t]
\centering
\includegraphics[scale=0.75]{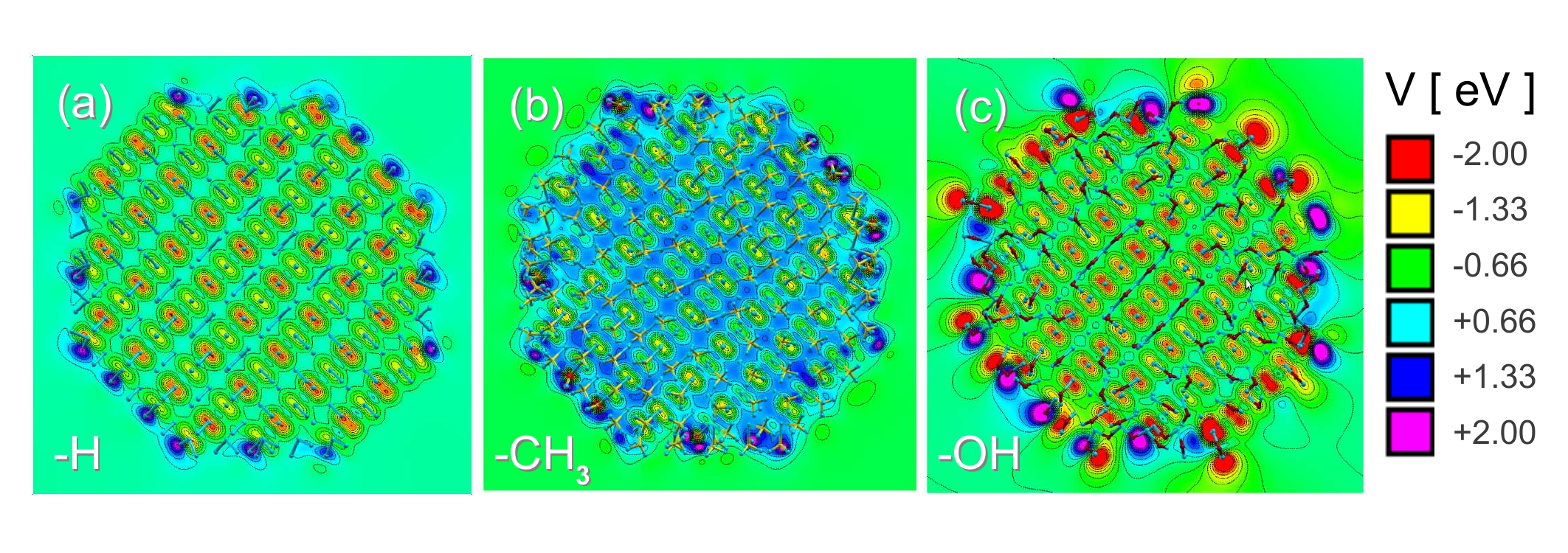}
\caption{ Map of Hatree potential inside Si$_{538}$ \nc s with different passivations. For non-polar passivating groups (a) --H and (b) --CH$_3$, the potential is regular and corresponds to the charge distribution in the crystal lattice. However, for polar --OH passivation (c), considerable local electrostatic fields can be seen near the surface. Note: Potential of neutral atoms was substracted from the full self-consistent potential to make the features easily visible. }
\label{Vhartree}
\end{figure*}

\begin{figure*}[!!!t]
\centering
\includegraphics[scale=0.5]{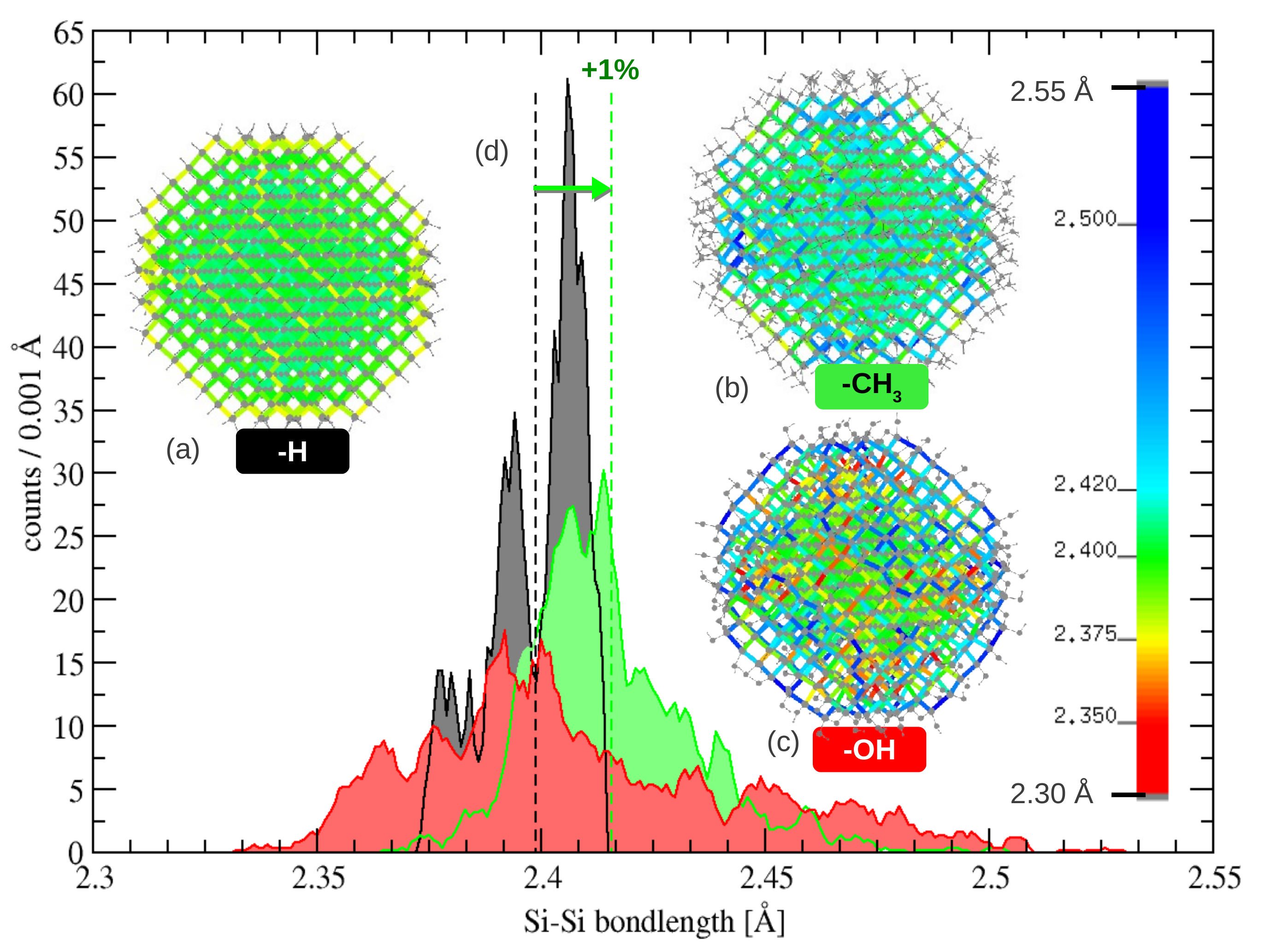}
\caption{ Histogram of Si-Si bond lengths in Si$_{538}$ \nc s with different passivation. Maps of bond length distortions are depicted as insets (a,b,c) in colorscale: (a) Si$_{538}$-H, (b) Si$_{538}$-CH$_3$ and (c) Si$_{538}$-OH. (d) Increase in the mean bond length between --H and --CH$_3$ by 1\%. }
\label{distortion_hist}
\end{figure*}

\subsection{SiNC:H}

First, we consider Si \nc s passivated by hydrogen groups (--H). Our calculations of fully optimized structure show that the atomic relaxation of Si atoms in the core region is negligible. Therefore, the Si \nc\ core has very similar atomic structure as bulk Si. We attribute this effect to a negligible charge transfer between the Si core and the hydrogen terminating groups due to fairly close electronegativities of both elements (1.9 for Si, 2.2 for H, respectively) and negligible mechanical stress resulting from a~small volume of the hydrogen terminating groups.

Calculated projected density of states (PDOS) for a Si$_{538}$ \nc\ is shown in \reffig{PDOS_gap}a. Surface states corresponding to Si-H bonds are localized far from the band gap edges. Therefore, highest occupied and lowest unoccupied electronic states are localized mainly in the core region. The electronic states in the Si \nc\ core are strongly affected by the quantum confinement effect. Consequently, the band gap of a Si \nc\ decreases with increasing \nc\ diameter (see \reffig{PDOS_gap}d), in good agreement with previous calculations \cite{Seino2010,Konig2008,Konig2009,Guerra2009}.

\reffig{k_and_real}a displays real space wave functions corresponding to individual electronic states on band gap edges. We see that,in case of hydrogen passivation, both HOMOs and LUMOs are spread almost homogeneously over the whole Si$_{538}$ core. Applying our transformation method, we can convert electronic states from real to reciprocal space to obtain the electronic band-like picture of a Si \nc\ with a given diameter. \reffig{k-lines-size} represents projected $\mathbf k$-space density along the $\Gamma-X$ direction. We can see the gradual emergence of the band-like structure with the increase of \nc\ size.  While for the smallest Si$_{68}$ \nc\ the band structure can be hardly recognized, in the case of the Si$_{538}$ \nc\ the dispersion of electronic states near the band gap  mimics well the bulk band structure of both the conduction and 
the valence bands (compare \reffig{k-lines-size}c,d). Therefore, we can estimate $\approx$2 nm to be a phenomenological limit where it makes sense to speak about electronic band structure and \textit{indirect} band-gap in silicon \nc s. 

There are two important differences between bulk band structure and projected band structure of a Si$_{538}$-H \nc\ that has to be noticed: (i) presence of discontinuities (mini-gaps) in energy dispersion and (ii) delocalization (blurring) of electronic states in $\mathbf k$-space. 

The presence of the mini-gaps within the bands is induced by the finite size of the Si \nc\ and the preserved symmetry of the atomic Si-core structure. The latter evokes the degeneracy of the electronic states with $\mathbf{k}$-vector in the equivalent crystal lattice directions in a similar manner as shown in \reffig{Symmetry_Breaking} for Si$_{68}$-H. We should note that the presence of mini-gaps larger the 63 meV can have important implications for the relaxation process of {\it hot} excited electrons from $\Gamma$ to $X$. Usually, the maximal vibrational energy of phonons in bulk Si does not exceed 63 meV \cite{Dargys1994}. Therefore {\it hot} electrons might be unable to reach the conduction band minimum near the $X$-points once they meet the mini-gap along their path from $\Gamma$ to $X$.

The second difference between the Si-bulk band structure and a hydrogen passivated \nc\ is the $\mathbf k$-space delocalization (blurring) of electronic states. This effect can be directly attributed to the confinement in a~finite space by window function $ w()$, as discussed in the~previous chapter. \reffig{k-delocalization_by_size} shows $\mathbf k$-space density projection of the lowest unoccupied state onto a 1-D line between the $\Gamma$(0,0,0) and $\Gamma$(0,2,2) $\mathbf k$-points for different Si:H \nc\ diameters. It demonstrates the variation of $\mathbf k$-space delocalization and its shape depending on the size of the Si \nc. We can clearly identify the $ \mathrm{sinc}$-like shape envelope corresponding to the rectangular step window function (see \reffig{H-chain-delocalization} for comparison). 

\subsection{SiNC:CH$_3$}

In the next step, we examine an~impact of methyl (--CH$_3$) terminating group on the atomic and electronic structure of Si \nc s.
Our fully relaxed Si:CH$_3$ \nc s expand by $\approx$ 1\% of lattice constant as a consequence of steric repulsion between the individual -CH$_3$ groups 
(see \reffig{distortion_hist}). The presence of mechanical strain causes inhomogeneities in the atomic Si-core structure, which lifts up the degeneracy of molecular states (see \reffig{k_and_real}b). This effect significantly reduces the size of the mini-gaps (see \reffig{k_and_real}b). Moreover, the electronic states are more localized in real space compared to those of a H-passivated \nc\ of a similar size as seen from comparison of \reffig{k_and_real}a,b. Consequently, the {\it tails} of projected densities $\tilde \rho_i(\mathbf{k'})$ of individual electronic states in $\mathbf{k}$-space decay more slowly in $\mathbf{k}$-space than in Si:H \nc s of the same size (compare again \reffig{k_and_real}a,b). This means that individual ``bands'' are not strictly localized in their $\mathbf k$-momentum.

We found out that the presence of non-polar methyl groups leads to the localization of surface states in energies far from the Si \nc\ band gap similarly as in Si:H \nc s (\reffig{PDOS_gap}b), but the band gap is smaller compared to Si:H \nc s (\reffig{PDOS_gap}d).

\subsection{SiNC:OH}

According to our DFT simulations, Si \nc s capped with hydroxyl (--OH) groups suffer from large atomic relaxation in proximity to the surface with respect to  Si bulk structure (see \reffig{distortion_hist}c). Nevertheless, atomic structure deep within the Si \nc\ core remains almost intact and thus close to the bulk. The large distortion of surface atomic structure is driven by the presence of significant charge transfer between Si and electronegative O atoms. The strong polarity of the Si-O bond tends to cause large localization of their MOs near the surface in real space (see \reffig{k_and_real}c) with energies close to band gap edges (\reffig{PDOS_gap}c). This picture is consistent with previous experimental evidence \cite{Wolkin1999PRL} and theoretical simulations \cite{Seino2010,Guerra2009,Guerra2010,Konig2008,Konig2009} of oxidized  Si \nc s. 

What's more, our simulations point out that --OH groups tend to align via hydrogen bonds, forming ordered domains at low temperature. Consequently, considerable local electric fields are induced across the Si \nc\ as shown in \reffig{Vhartree}. We computed the electric dipole of 120 Debye of the largest --OH passivated \nc\ (3 nm, Si$_{538}$), while the electric dipoles of the corresponding \nc s passivated by --H (0.15 Debye) and --CH$_3$ (0.9 Debye) are negligible. The presence of these irregularities in electric field together with the geometrical distortion of the surface layer (\reffig{distortion_hist}c) probably cause the irregular real space distribution of HOMOs and LUMOs in --OH passivated \nc s (\reffig{k_and_real}c).   

Due to the presence of strongly localized states in the surface layer, the projected band structure is significantly blurred (with large \textit{tails} of projected density $\tilde \rho_i(\mathbf{k'})$), which relaxes $\mathbf k$-space selection rules of optical transitions and improves the radiative recombination probability of {\it indirect} transitions. Also, the degeneracy of electronic states is almost removed due to the large distortion of atomic structure near the surface. This, however, goes hand in hand with higher localization and irregular distribution of MOs in real space, which decrease real space overlap of orbitals and thus can limit the radiative recombination probability.

\section{Discussion}

In this article, we show that band structure can be rigorously defined in nanostructures, taking into account several alterations with respect to bulk materials. The first difference is that the utilization of the band-structure concept makes sense only for ``larger'' nanocrystals, with sizes above a certain size limit. For example, for Si \nc s this size limit lies between 1.5 and 2 nm (see \reffig{k-lines-size}), because under this limit the $\mathbf{k}$-space projected MOs do not show much of band-like behavior. 

Second, the smaller the nanocrystal, the more pronounced the presence of mini-gaps inside the energy bands, which can limit the non-radiative relaxation of excitons. For the same size nanocrystals, the mini-gaps are wider in more symmetric nanocrystals with weaker effect of surface due to the degeneracy of energy levels (see \reffig{k_and_real}).  

Third, the nanoscale size introduces blurring of the $\mathbf k$ vector (see \reffig{k-delocalization_by_size} and \reffig{k-convolution}d). Although this phenomenon is qualitatively easily predictable from the Heisenberg uncertainty principle and was already treated e.g.\ by Hybertsen,\cite{HybertsenPRL1994} our approach allows us to quantify the influence of $\mathbf k$ vector blurring for  individual MOs. 

Last, in addition to describing the crystalline core, our computed band structures already include the effect of surface states. This is very important, because many types of real-life semiconductor nanocrystals need to be capped by various surface-terminating groups, which profoundly influences their electronic properties.

\section{Conclusion}
In this paper, we introduced  the general method to map effectively electronic structure of aperiodic systems such as \nc s from the real space to reciprocal space.  
This method allows us to reassemble the electronic band structure of finite size systems. We believe that this method could provide more insight into the question if the band structure concept can 
be still applied to nanocrystals of different shape, size or chemical composition. In particular, we demonstrate that the band structure picture of nm-scale Si \nc s can be still adopted down to $\approx$2 nm size, but with two important consequences of the finite size of system:  (i) the discretization (ii) the delocalization of electronic states in the reciprocal space. This results also means that efficient slow red photoluminescence in Si \nc s arises from indirect X-to-$\Gamma$ electron-hole recombination, in agreement with recent experimental evidence \cite{Hannah2012}. 

We employed this method to investigate the effect of Si \nc\ size and presence of different passivation groups including non-polar and polar groups on their band structure.  
We found, that Si \nc s caped with methyl group expand by $\sim$1 $\%$ due to steric repulsion between methyl surface groups.  
In case of polar hydroxyl group, real space localized states near surface form band gap edges, which tends toward rather delocalized $\mathbf{k}$-space states. In other words, 
the band structure preserves, but near the band gap the electronic states becomes more blurred. In addition, we found strong alignment of hydroxyl group forming  significant 
macroscopic electrostatic dipole moment across Si \nc.

\section{ Acknowledgement }

This research was supported by GACR 202/09/H041 project and GACR Centrum excelence P108/12/G108. We also acknowledge Jan Valenta for many useful discussions.

\end{document}